# Cellular automaton model with dynamical 2D speed-gap relation reproduces empirical and experimental features of traffic flow

Junfang Tian

Institute of Systems Engineering, College of Management and Economics, Tianjin University, No. 92 Weijin Road, Nankai District, Tianjin 300072, China, jftian@tju.edu.cn

Bin Jia

MOE Key Laboratory for Urban Transportation Complex Systems Theory and Technology, Beijing Jiaotong University, Beijing, China, bjia@bjtu.edu.cn

Shoufeng Ma, Chenqiang Zhu

Institute of Systems Engineering, College of Management and Economics, Tianjin University, No. 92 Weijin Road, Nankai District, Tianjin 300072, China, {sfma@tju.edu.cn, sandyzhucn@126.com}

Rui Jiang, YaoXian Ding

MOE Key Laboratory for Urban Transportation Complex Systems Theory and Technology, Beijing Jiaotong University, Beijing, China, {rjiang@ustc.edu.cn, dingyaoxian@gmail.com}

This paper proposes an improved cellular automaton traffic flow model based on the brake light model, which takes into account that the desired time gap of vehicles is remarkably larger than one second. Although the hypothetical steady state of vehicles in the deterministic limit corresponds to a unique relationship between speeds and gaps in the proposed model, the traffic states of vehicles dynamically span a two-dimensional region in the plane of speed versus gap, due to the various randomizations. It is shown that the model is able to well reproduce (i) the free flow, synchronized flow, jam as well as the transitions among the three phases; (ii) the evolution features of disturbances and the spatiotemporal patterns in a car-following platoon; (iii) the empirical time series of traffic speed obtained from NGSIM data. Therefore, we argue that a model can potentially reproduce the empirical and experimental features of traffic flow, provided that the traffic states are able to dynamically span a 2D speed-gap region.

# 1. Introduction

Empirical investigations show that the congested traffic exhibits very complex spatiotemporal phenomena, such as the capacity drop, phantom jam, widely scattered data on the flow-density plane and lane changing (see: Chowdhury, 2000; Helbing, 2001; Nagatani, 2002; Kerner, 2004, 2009; Laval, 2011; Treiber and Kesting, 2013; Sun and Elefteriadou, 2012; Jin, 2014; Zheng, 2014). In order to explore these phenomena, a huge number of traffic flow models were proposed, which have been classified by Kerner into the two-phase models (see Helbing, 2001; Nagel et al., 2003; Treiber and Kesting, 2013) and the three-phase models (see Kerner and Rehborn 1996a, 1996b, 1997; Kerner, 1998, 2004, 2009;).

Two-phase models usually presume that there is a unique relationship between the flow rate and the traffic density under the steady state condition, which is called fundamental diagram. In these models, the traffic flow is classified into the Free Flow (FF) phase and congested flow phase. Based on a long-term empirical analysis, Kerner introduced the Three-Phase Theory (KTPT), which classifies the congested traffic into the Synchronized Flow (SF) and the Wide Moving Jams (WMJs). The fundamental difference between these two phases is that the downstream fronts of WMJs propagate upstream through the highway and bottlenecks with a characteristic mean velocity, while the downstream fronts of SF do not exhibit this characteristic feature and are usually fixed at the bottleneck. Usually one observes the phase transitions from FF to SF (F→S transition) and from SF to WMJs (S→J transition). KTPT believes that in general situations, WMJs can emerge only in SF and F→J transition cannot happen. F→J transition can occur only if the formation of SF is strongly hindered due to a non-homogeneity, in particular at a traffic split on a highway (Kerner, 2000). In contrast, phase transition involved in two-phase models is F→J transition. KTPT assumes that the hypothetical steady states of SF cover a two-dimensional region in the flow-density plane. In order to improve the readability, we have made an abbreviations list in Table A1 in the appendix.

Kerner (2009) claimed that the methodology of transportation engineering based on the two-phase traffic flow models is inconsistent with spatiotemporally measured traffic data and therefore the methodology leads to traffic flow theories that cannot be used for efficient dynamic management and control in congested freeway networks. Therefore KTPT was introduced to explain traffic breakdown at highway bottlenecks. Models within KTPT incorporate the fundamental hypothesis that the hypothetical steady states of SF should cover a two dimensional region in the flow-density plane. Kerner (2012) states that macroscopic and microscopic spatiotemporal effects of the entire complexity of traffic congestion observed up to now in real measured traffic data can be explained by the simulations of traffic flow consisting of identical drivers and vehicles, if a microscopic model used in these simulations incorporates the fundamental hypothesis of KTPT.

We would like to mention that Kerner (2009) also pointed out that the link between three-phase traffic theory and the FD of traffic flow can be created through the use of the averaging of an infinite number of steady states of SF into one SF for each density. In this case, these three-phase models can reproduce some features of real traffic, see e.g., Speed-Adaptation Models (SAMs) proposed by Kerner and Klenov (2006). However, they will lose the possibility of the description of very important features of SF found in empirical observations. For instance, SAMs have been criticized for the disability to reproduce the observed LSPs as well as some empirical features of SF between WMJs within General Patterns (GPs) (Kerner and Klenov, 2006).This emphasizes the sense and importance of 2D steady states of SF for the development of three-phase traffic flow models.

It is not surprised that there are many controversies between the two-phase theory and KTPT (see Schönhof and Helbing, 2007, 2009; Helbing et al. 2009; Treiber et al. 2010; Daganzo et al. 1999; Kerner, 2004, 2009, 2012, 2013; Kerner and Klenov, 2002, 2006; Kerner et al. 2004, 2011, 2014). The two-phase models have been criticized for not able to describe the empirical features of traffic breakdown as well as further developments of related congested regions properly while models within KTPT are criticized for their complexity and having too many model parameters. In particular, recently Jin et al. (2013) have found an empirical F→S transition measured on a single-lane highway in China that can be simulated by models within KTPT, but two-phase models fail to reproduce. Jiang et al. (2014)

carried out a large-scale car-following experiment on an open road section, which shows that the nature of car-following runs against two-phase models such as optimal velocity model (Bando et al., 1995), full velocity difference model (Jiang et al., 2001), and intelligent driver model (Treiber et al., 2000), but is consistent with KTPT. As agreed in the traffic flow community, the best approach to solve these ongoing controversies is to collect more precise traffic data via empirical investigations and traffic flow experiments.

In this paper, an improved cellular automaton model based on the Brake Light Model (abbreviated as BLM, Knospe et al, 2000) is proposed. BLM is one of the most popular models, which has been used by Hafstein et al. (2004) to simulate traffic flow on the autobahn network in North Rhine-Westphalia; by Mallikarjuna and Rao (2009) to study heterogeneous traffic observed in developing countries; by Appert-Rolland and Boisberranger (2011) to test lane changing rules; by Meng and Weng (2011) to simulate the heterogeneous traffic in work zone; and by Knorr et al. (2012) to analyze the strategy to reduce traffic congestion with the help of vehicle-to-vehicle communication. However, BLM has been criticized by KTPT. Kerner et al. (2004) pointed out that its simulation results are inconsistent with the empirical observations. At the same given flow rates on the one-lane main road and on the on-ramp, the Oscillating Moving Jams (OMJ) is formed in BLM while the Widening SF Pattern (WSP) is reproduced by KKW within KTPT (Kerner et al.,2004). OMJ is characterized by a complex birth and decay of narrow moving jams while no moving jam emerges in WSP. In KKW model, if the on-ramp flow is further gradually increased, WSP spontaneously transforms into either Dissolving General Pattern (DGP) or GP, and GP will not change into other congested patterns with the further increase of the on-ramp flow. In contrast, if the flow rate on the on-ramp increases in BLM, the Widening Pinned Layer (WPL) develops, whose downstream front is pinned at the on-ramp and upstream front is slowly moving upstream. The mean vehicle speed and flow rate in WPL is very low, about 10*km/h* and 480*veh/h* respectively. In the most upstream of WPL, the OMJ occurs spontaneously while SF is realized in the upstream of WMJs in GP.

In the proposed model in the present paper, although the hypothetical steady state of vehicles corresponds to a unique relationship between speeds and gaps, the traffic state of vehicles dynamically spans a two-dimensional region in the plane of speed versus gap, due to the various randomizations. It is shown that the model is able to well reproduce (i) the FF, SF, WMJ as well as the transitions among the three phases; (ii) the evolution features of disturbances and the spatiotemporal patterns in a car-following platoon; (iii) the empirical time series of traffic speed obtained from NGSIM data. Therefore, we argue that a model can potentially reproduce the empirical and experimental features of traffic flow, provided that the traffic states are able to dynamically span a 2D speed-gap region.

The paper is organized as follows. Section 2 analyzes the deficiency of BLM and Section 3 establishes the new model. In Section 4, we carry out numerical simulations, which show that the new model can reproduce the phase transitions and the experimental spatiotemporal dynamics of traffic flow. Section 5 calibrates and validates the new model by the empirical data. Section 6 concludes the paper.

## 2 BLM analysis

BLM combines the velocity anticipation and slow-to-start rule into the NaSch model (Nagel and Schreckenberg, 1992). The most important contribution is that a dynamical long ranged interaction was introduced: drivers react on the braking of the leading vehicle indicated by the brake light. The parallel update rules of BLM are consisted by the following five steps.

**Step 1. Determination of the randomization parameter** $p$**:**

$$p = p(v_n(t), b_{n+1}(t), t_{n,h}(t), t_{n,sa}(t)) = \begin{cases} p_b : & \text{if } b_{n+1} = 1 \text{ and } t_{n,h} < t_{n,sa} \\ p_0 : & \text{if } v_n = 0 \\ p_d : & \text{in all other cases} \end{cases}$$

$b_n(t+1) = 0$

**Step 2. Acceleration:**

if $((b_{n+1}(t) = 0 \text{ and } b_n(t) = 0) \text{ or } t_{n,h} \geq t_{n,sa})$ then

$v_n(t+1) = \min(v_n(t) + a_1, v_{max})$

else $v_n(t+1) = v_n(t)$

**Step 3. Braking rule:**

$v_n(t+1) = \min(v_n(t+1), d_{n,eff}(t))$

if $(v_n(t+1) < v_n(t))$ then $b_n(t+1) = 1$

**Step 4. Randomization and braking:**

if $(rand() < p)$ then $\begin{cases} v_n(t+1) = \max(v_n(t+1) - d_1, 0) \\ \text{if } (p = p_b) \text{ then } b_n(t+1) = 1 \end{cases}$

**Step 5. Car motion:**

$x_n(t+1) = x_n(t) + v_n(t+1)$

where $t$ is the time step; $d_n = x_{n+1} - x_n - L_{veh}$ is the gap between vehicle $n$ and its preceding vehicle $n+1$; $L_{veh}$ is the length of vehicles; $x_n$ and $v_n$ are the position and speed of vehicle $n$; $b_n$ is the status of the brake light ($b_n = 1(0)$ means the brake lights is on (off)); $t_{n,h} = d_n/v_n$ is the time gap that vehicle $n$ reaches the position of the rear of vehicle $n+1$; $t_{n,sa} = \min(v_n, h)$ is the safe time gap of vehicle $n$ and $h$ determines the range of interaction with the brake light; $d_{n,eff} = d_n + \max(v_{anti} - g, 0)$ is the effective gap; $v_{anti} = \min(d_{n+1}, v_{n+1})$ is the expected speed of the preceding vehicle in the next time step; $g$ is the parameter to control the effectiveness of anticipation and keep safety and accidents are avoided only if the constraint $g \geq d_1$ is satisfied; $rand()$ is a random number between 0 and 1; $p_b$, $p_0$, and $p_d$ are the randomization probabilities; the acceleration capacity is assumed to be $a_1$; the randomization deceleration capacity is assumed to be $d_1$. $d_1 \geq a_1$ should be satisfied to simulate WMJs. The values of all parameters are adopted as that in Knospe et al. (2000).

Fig.1 is a part of vehicle' trajectories of BLM under periodic boundary condition at the density 27*veh/km*. Trajectories of vehicle 1 and 2 show that there is a narrow moving jam. To analyze the emergence mechanism of this jam, the states of them are listed in Table 1. From the variation tendency of the time gap $t_{2,h}$, it can be concluded that the zero speeds of vehicle 2 are caused by the small time gaps. When the leading vehicle decelerates sharply at $t=30200s$, there is no enough time for vehicle 2 to adapt its speed smoothly due to the reaction delay. Thus, the braking rule of BLM is revised as follows:

**Step 3. Revised braking rule:**

$v_n(t+1) = \min(v_n(t+1), \lceil d_{n,eff}(t)/T \rceil)$

if $(v_n(t+1) < v_n(t))$ then $b_n(t+1) = 1$

where $\lceil x \rceil$ is the minimum integer that bigger than $x$; $T > 1$ is the desire time gap that vehicles hope to keep. Compared with $T = 1$ in the BLM model, this reflects the fact that cars tend to moving more slowly in order to avoid unrealistic oversized deceleration in the next time step. Fig.2 shows the spatiotemporal diagram at the density 27*veh/km* by the BLM with the revised braking rule. The speeds of vehicle 2 are no longer zero, thus the slow-to-start rule will not be triggered, and narrow moving jams will not occur.

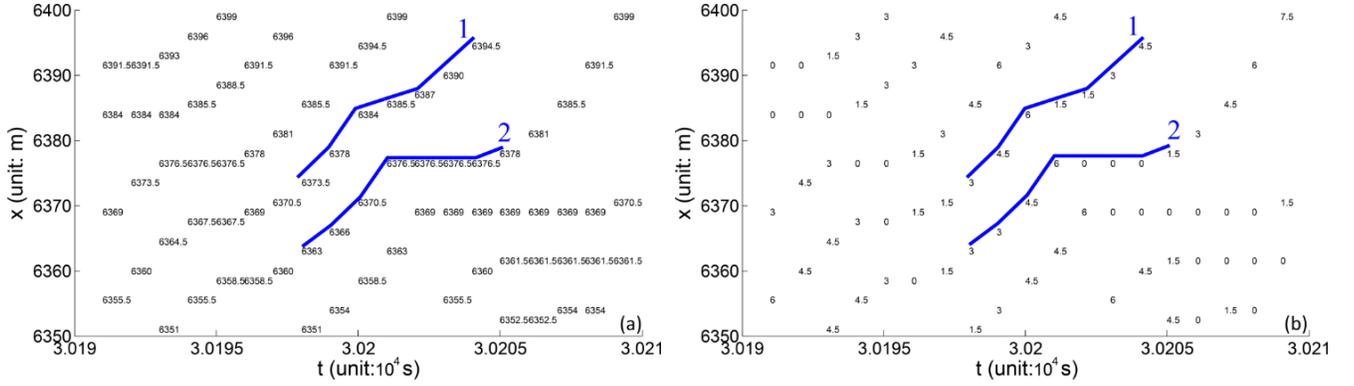

**Fig. 1.** Part of vehicle' trajectories of BLM under periodic boundary condition at the density 27*veh/km*. (a, b) is vehicles' positions (unit: *m*) and speeds (unit: *m/s*) respectively. The horizontal direction (from left to right) is time and the vertical direction (from down to up) is space. The parameter values of BLM are taken from Knospe et al. (2000).

**Table 1.** The states of vehicle 1 and 2. $b_1$, $x_1$ and $v_1$ are the brake light state, position and speed of vehicle 1; $x_2$, $d_2$ $v_2$, and $t_{2,h}$ are the position, space gap, speed and time gap of vehicle 2.

| Time (unit:*s*) | 30198 | 30199 | 30200 | 30201 | 30202 | 30203 | 302004 |
|---|---|---|---|---|---|---|---|
| $b_1$ | 0 | 0 | 0 | 1 | 0 | 0 | 0 |
| $x_1$(unit: *m*) | 6373.5 | 6378 | 6384 | 6385.5 | 6387 | 6390 | 6394.5 |
| $v_1$(unit: *m/s*) | 3 | 4.5 | 6 | 1.5 | 1.5 | 3 | 4.5 |
| $x_2$ (unit: *m*) | 6363 | 6366 | 6370.5 | 6376.5 | 6376.5 | 6376.5 | 6378 |
| $d_2$ (unit: *m*) | 3 | 4.5 | 6 | 1.5 | 3 | 6 | 9 |
| $v_2$ (unit: *m/s*) | 3 | 3 | 4.5 | 6 | 0 | 0 | 0 |
| $t_{2,h}$ (unit: *s*) | 1 | 1.5 | 1.33 | 0.25 | - | - | - |

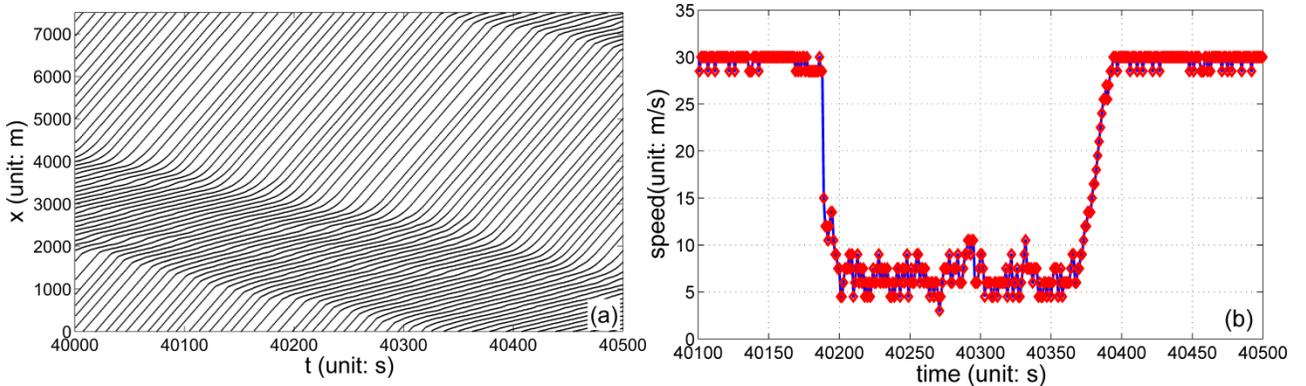

**Fig. 2.** (a) The spatiotemporal diagram of every 5th vehicle after introducing the desire time gap *T*=1.8*s* at the density 27*veh/km*. (b) the speed series of one of the vehicles taken from (a).

## 3. The new model

Based on the above analysis, an improved model is proposed with the following parallel update rules:

**Step 1. Determining the randomization probability *p*:**

$$p = p(v_n(t), b_{n+1}(t), t_{n,h}(t), t_{n,sa}(t)) = \begin{cases} p_b: & \text{if } b_{n+1} = 1 \text{ and } t_{n,h} < t_{n,sa} \\ p_0: & \text{if } v_n = 0 \\ p_d: & \text{in all other cases} \end{cases}$$

$b_n(t+1) = 0$

**Step 2. Acceleration:**

if $((b_{n+1}(t) = 0$ or $t_{n,h}(t) \geq t_{n,sa}(t))$ and $v_n(t) > 0)$ then

$v_n(t+1) = \min(v_n(t) + a_1, v_{max})$

else $v_n(t+1) = \min(v_n(t) + a_2, v_{max})$

**Step 3. Braking rule:**

$v_n(t+1) = \min(v_n(t+1), \lceil d_{n,eff}(t)/T \rceil)$

if $(v_n(t+1) < v_n(t))$ then $b_n(t+1) = 1$

**Step 4. Randomization and braking:**

if $(rand() < p)$ then $\begin{cases} v_n(t+1) = \max(v_n(t+1) - d_1, 0) \\ \text{if } (p = p_b) \text{ then } b_n(t+1) = 1 \end{cases}$

**Step 5. Car motion:**

$x_n(t+1) = x_n(t) + v_n(t+1)$

Note that in step 2, two acceleration abilities $a_1$ and $a_2$ have been adopted for vehicles in different state, which may either equal to each other or not. Compared with BLM, the highlight feature is that the desire time gap is set as $T > 1$. Thus, this model is named as the Desire Time Gap BLM, abbreviated as DTGBLM. Although the difference seems trivial, the performances of DTGBLM are far better than that of BLM, see the following sections.

Next, the steady states of DTGBLM are considered. In the deterministic limit $p_b = 1$, $p_d = 0$, and $p_0 = 0$, the steady states are given by $v_n = \min(d_{n,eff}/T, v_{max})$. Thus the unique relationship between speed and space gap is established.

Nevertheless, in traffic flow, the stochastic factors play an important role. Fig.3(a, b) show two examples of the velocities of a car (which is asked to move with constant speed) and its following car, as well as the gap between the two cars in Jiang et al.'s traffic experiment (Jiang et al., 2014). One can see that while the speed of the following car fluctuates weakly, the gap changes significantly. This demonstrates that the steady states do not exist in real traffic and the traffic state actually dynamically spans a two dimensional speed-gap region.

In the DTGBLM, when the various randomizations are considered, the traffic state is able to dynamically span a two dimensional speed-gap region as in the experiment, see Fig.3(c, d). The model could well reproduce the empirical and experimental features of traffic flow as shown in sections 4 and 5.

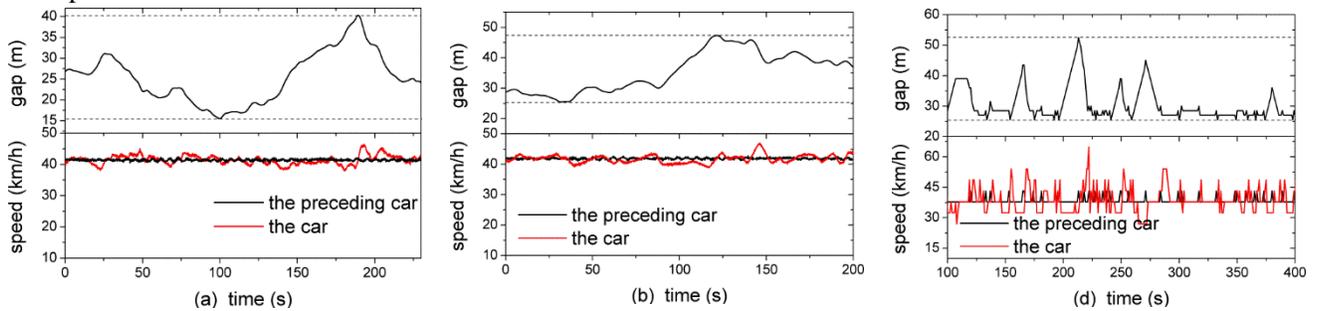

**Fig. 3.** (a, b) The evolution of speeds of a car and its preceding car and their gaps taken from the real car following experiments. The preceding car in (a) is the same as that of (b), while their following cars are different. (c, d) Simulation results with the DTGBLM with the parameters in Table 2. During the simulations, the leading car moves with constant speed but subject to randomization with probability (c) $p = 0.5$, (d) $p = 0.9$ to activate the brake light related randomization rule.

## 4 Simulation analysis

Firstly, we simulate traffic flow on a circular road and on an open road with an on-ramp as one normally tests a traffic flow model. Next, car-following behaviors have been simulated to test whether DTGBLM can reproduce the experimental evolution feature of disturbances and the spatiotemporal patterns as revealed in the car-following experiment (Jiang et al., 2014).

Simulations are carried out on a road with the length $L_{road}= 1000L_{veh}$. The cell length and vehicle length are set as 1.5*m* and 7.5*m*, respectively, i.e. $L_{cell}= 1.5m$ and $L_{veh} = 5L_{cell}= 7.5m$. One time step corresponds to 1*s*. The parameter values are shown in Table 2, which are regarded as default values and most of which are the same as that of BLM.

**Table 2.** Default parameters of DTGBLM.

| Parameters | $L_{cell}$ | $L_{veh}$ | $v_{max}$ | $h$ | $T$ | $p_b$ | $p_0$ | $p_d$ | $g$ | $a_1$ | $a_2$ | $d_1$ |
|---|---|---|---|---|---|---|---|---|---|---|---|---|
| Units | *m* | $L_{cell}$ | $L_{cell}/s$ | *s* | *s* | - | - | - | $L_{cell}$ | $L_{cell}/s$ | $L_{cell}/s$ | $L_{cell}/s$ |
| Value | 1.5 | 5 | 20 | 6 | 1.8 | 0.94 | 0.5 | 0.1 | 7 | 2 | 1 | 1 |

*4.1 Circular road*

As usual, the following two initial configurations are used in the simulations: 1) all vehicles are homogeneously distributed on the road; 2) all vehicles are distributed in a megajam. Fig.4 is the flow-density diagram of DTGBLM, where the FF and SF branches are from the initial homogeneous distribution, while the WMJ branch is from initial megajam. Fig.5 shows the F→S and S→J transitions which demonstrate that the DTGBLM can well reproduce the FF, the SF, and the jams, as well as phase transitions among the three phases.

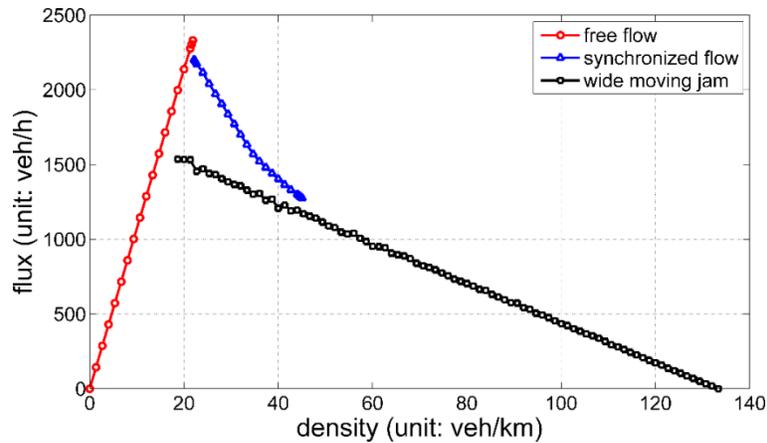

**Fig. 4.** Flow-density diagram of DTGBLM.

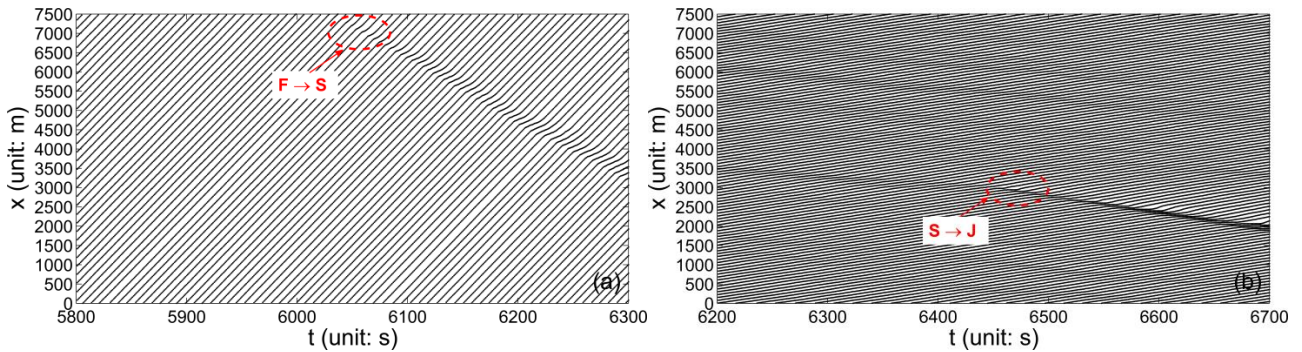

**Fig. 5.** The F→S and S→J transitions of DTGBLM on the circular road from a homogenous initial distribution. (a) the density $k = 22 veh/km$; (b) $k = 67 veh/km$. The horizontal direction (from left to right) is time and the vertical direction (from down to up) is space.

Fig.6 is the spatiotemporal diagrams of DTGBLM on the circular road: (a, b) fall on the SF branch while (c, d) locate on the WMJ branch. (a, b) show that when the density is small, traffic states represent the coexistence of FF and SF on the SF branch; with the increase of the density, FF will eventually disappear and only SF can be found. (c, d) show the coexistence of FF and WMJ, which shows the same variation trend as that of the SF branch. Fig.7 is the vehicles' speed series taken from Fig.6(a, c). The SF of DTGBLM contains no narrow moving jams, since the SF speed lies in the range of [9, 25]$m/s$ on Fig.7(a). Moreover, there are some nonzero speed in the WMJ on Fig.7(b), which is more consistent with the real traffic flow compared with other cellular automata models, see Kerner and Rehborn (1996a, 1996b).

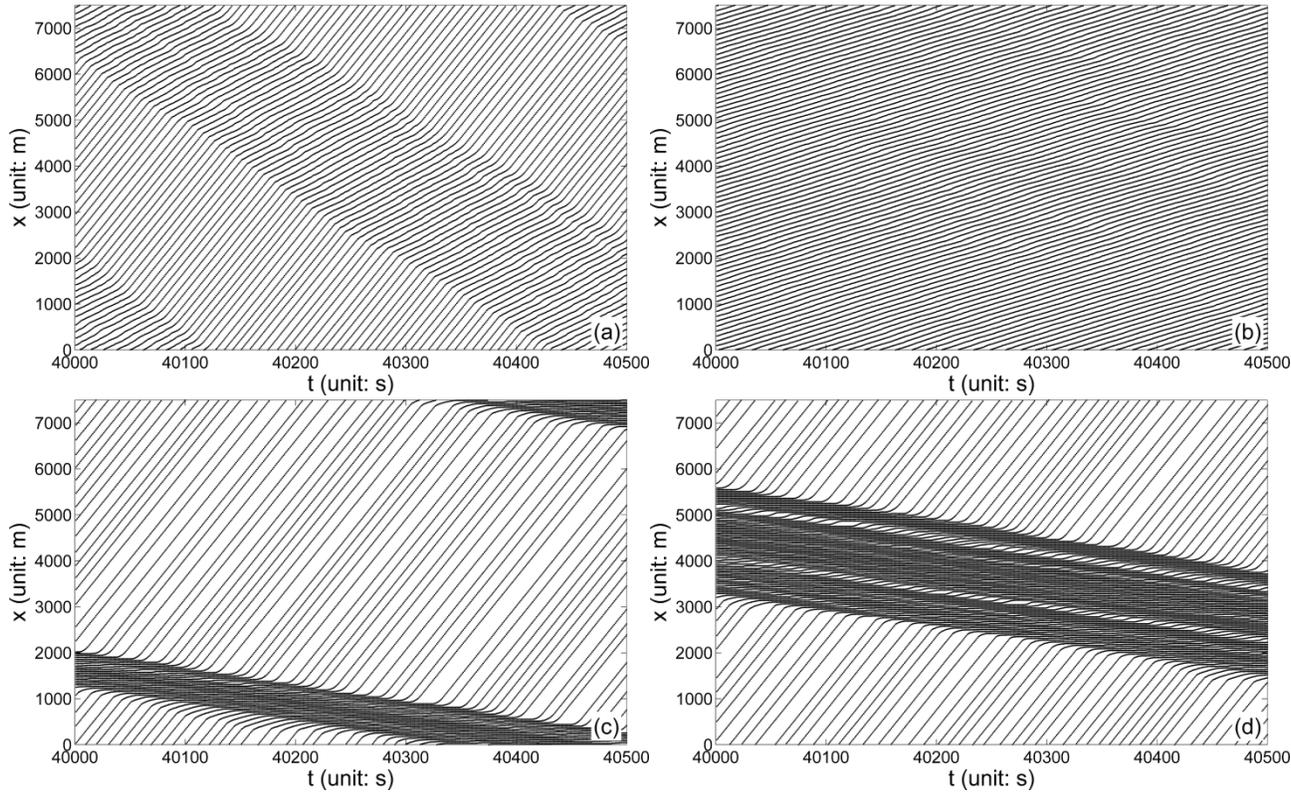

**Fig. 6.** The spatiotemporal diagrams of DTGBLM on the circular road. (a, c) the density $k=27veh/km$; (b, d) $k=40veh/km$. In (a, b) the traffic starts from a homogenous initial distribution. In (c, d) the traffic starts from a megajam. The horizontal direction (from left to right) is time and the vertical direction (from down to up) is space.

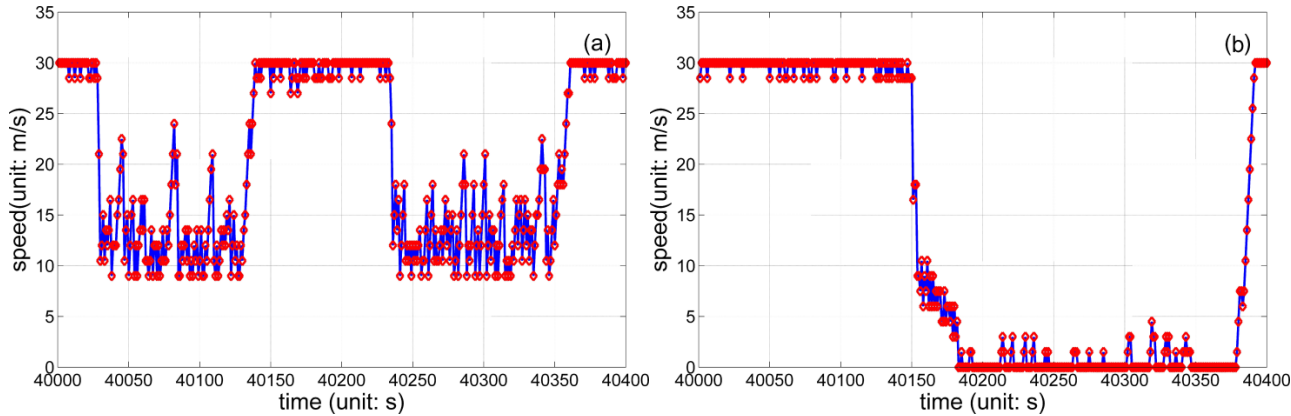

**Fig. 7.** The vehicle's speed series of one of the vehicles in DTGBLM. (a, b) corresponds to Fig.6(a, c) respectively.

*4.2 Open road with an on-ramp*

The traffic patterns that emerge near an on-ramp are studied on an open road. Vehicles drive from left to right. The left-most cell corresponds to $x=1$. The position of the left-most vehicle is $x_{\text{last}}$ and that of the right-most vehicle is $x_{\text{lead}}$. At each time step, if $x_{\text{last}}>v_{\max}$, a new vehicle with speed $v_{\max}$ will be injected to the position $\min(x_{\text{last}}-v_{\max}, v_{\max})$ with probability $q_{\text{in}}/3600$, where $q_{\text{in}}$ is the traffic flow entering the main road in units of vehicles per hour. At the right boundary, the leading vehicle moves without any hindrance. If $x_{\text{lead}}>L_{\text{road}}$, the leading vehicle will be removed and the following vehicle becomes the leader.

The method proposed by Kerner et al. (2002) is adopted to model the on-ramp. Assuming the position of the on-ramp is $x_{\text{on}}$, a region $[x_{\text{on}}, x_{\text{on}}+L_{\text{ramp}}]$ is selected as the inserting area of the vehicle from on-ramp. At each time step, two consecutive vehicles on the main road within the on-ramp area are chosen randomly and their coordinates are denoted by $x_l$ and $x_f$. The entering vehicle will be placed at the coordinate $x_{\text{ent}} = \lfloor x_f + L_{\text{veh}} + (x_l - x_f - 2L_{\text{veh}})/2 \rfloor$ (where $\lfloor x \rfloor$ denotes the biggest integer smaller than $x$) with probability $q_{\text{on}}/3600$ and $q_{\text{on}}$ is the traffic flow from the on-ramp, if the space gap $(x_l - x_f - L_{\text{veh}})$ between the two vehicles on the main road exceed the value $d_{\text{cri}}$ $(=\alpha v_l + \beta L_{\text{veh}})$. The speed of the inserted vehicle is set as $v_l$, which is the speed of the leading vehicle. The parameters are set as $x_{\text{on}}=0.8L_{\text{road}}$, $L_{\text{ramp}}=10L_{\text{veh}}$, $\alpha=0.55$ and $\beta=1.3$.

Fig.8 is the spatiotemporal diagrams of DTGBLM under open boundary condition with on-ramp. One can see that the WSP and GP can be reproduced by DTGBLM, which are totally different from that of BLM (see Fig. 21-24 in Kerner et al. (2002)).

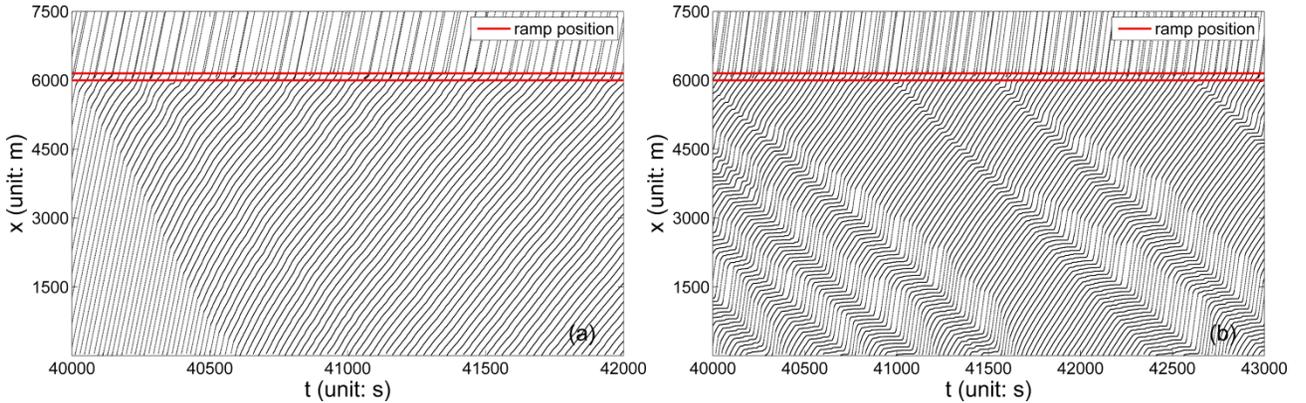

**Fig. 8.** The spatiotemporal diagrams of DTGBLM model on an open road with an on-ramp. (a, b) are the WSP and GP reproduced by DTGBLM, respectively.

*4.3 Car-following behaviors*

Jiang et al. (2014) carried out the controlled car following experiments concerning a platoon of 25 passenger cars on a 3.2-km-long open road section. The leading vehicle was asked to move with different constant speed. They found that when the velocity of the leading car is low, the standard deviation increase linearly with car number. With the increase of the velocity of the leading car, the standard deviation increase in a concave way, see Fig.9. Through the comparisons with the simulation results of car following models, they concluded that the nature of car-following runs against the traditional traffic flow theory. Simulations show that by removing the fundamental notion (i.e. the FD) in the traditional car-following models and allowing the traffic state to span a two-dimensional region in velocity-spacing plane, the growth pattern of disturbances has changed and becomes qualitatively or even quantitatively in consistent with that observed in the experiment.

Fig.9 simulates Jiang's experiment for DTGBLM. The jam initial condition is adopted and the leading vehicle

(No. 1) moves with the constant speed $v_l$. It can be seen that the disturbance evolution process of DTGBLM is in pretty good agreement with that of the experiment. Fig.10 compares the spatiotemporal patterns of the speeds of the car-platoon. One can see that the simulation results are very similar to the experimental ones.

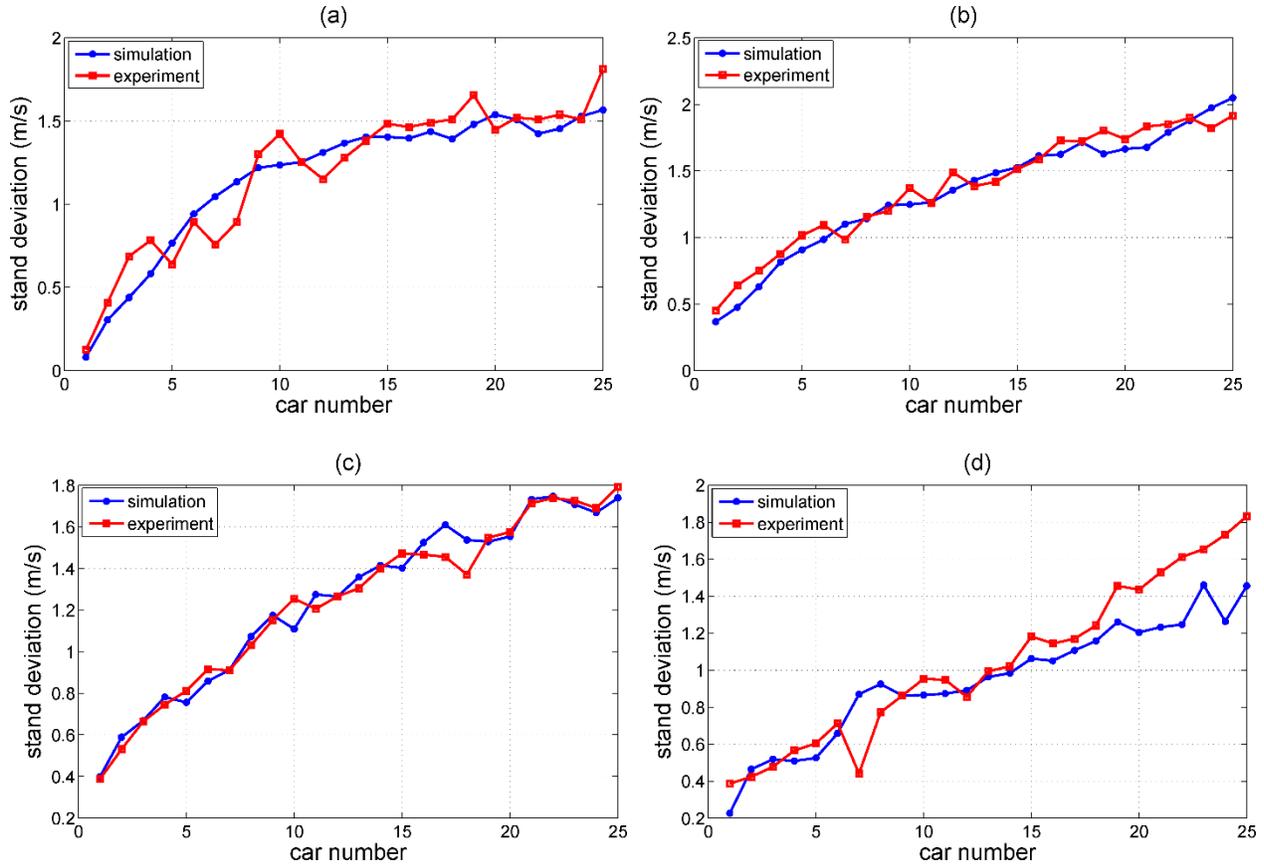

**Fig. 9.** The standard deviation of the time series of the speed of each car. The car number 1 is the leading car. In (a-d), the experimental leading vehicle was asked to move with the constant speed $v_{l,exp}$ =50, 40, 30 and 15$km/h$ respectively. The simulation results of DTGBLM are obtained with parameter values: $L_{cell}$= 0.5$m$, $L_{veh}$ =15$L_{cell}/s$, $a_1$=1$L_{cell}/s$, $v_{max}$ = 45$L_{cell}/s$ and $p_d$=0.3. Other parameters are the same as in Table 2. In the simulation, the velocity of the leading car is set by: $v_l$= [$v_{l,exp}/L_{cell}$], where [$x$] is the integer nearest to $x$.

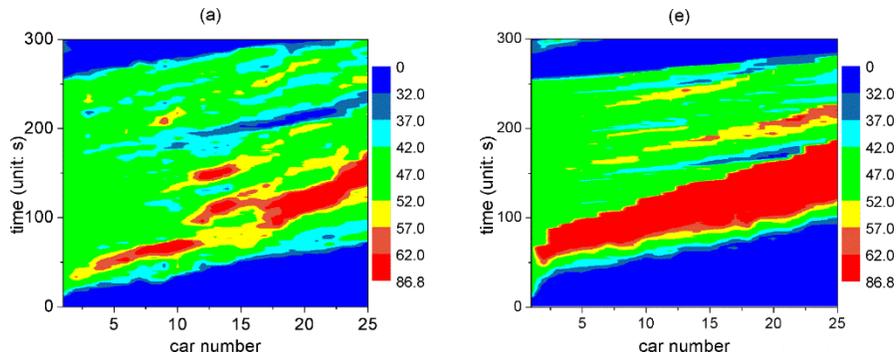

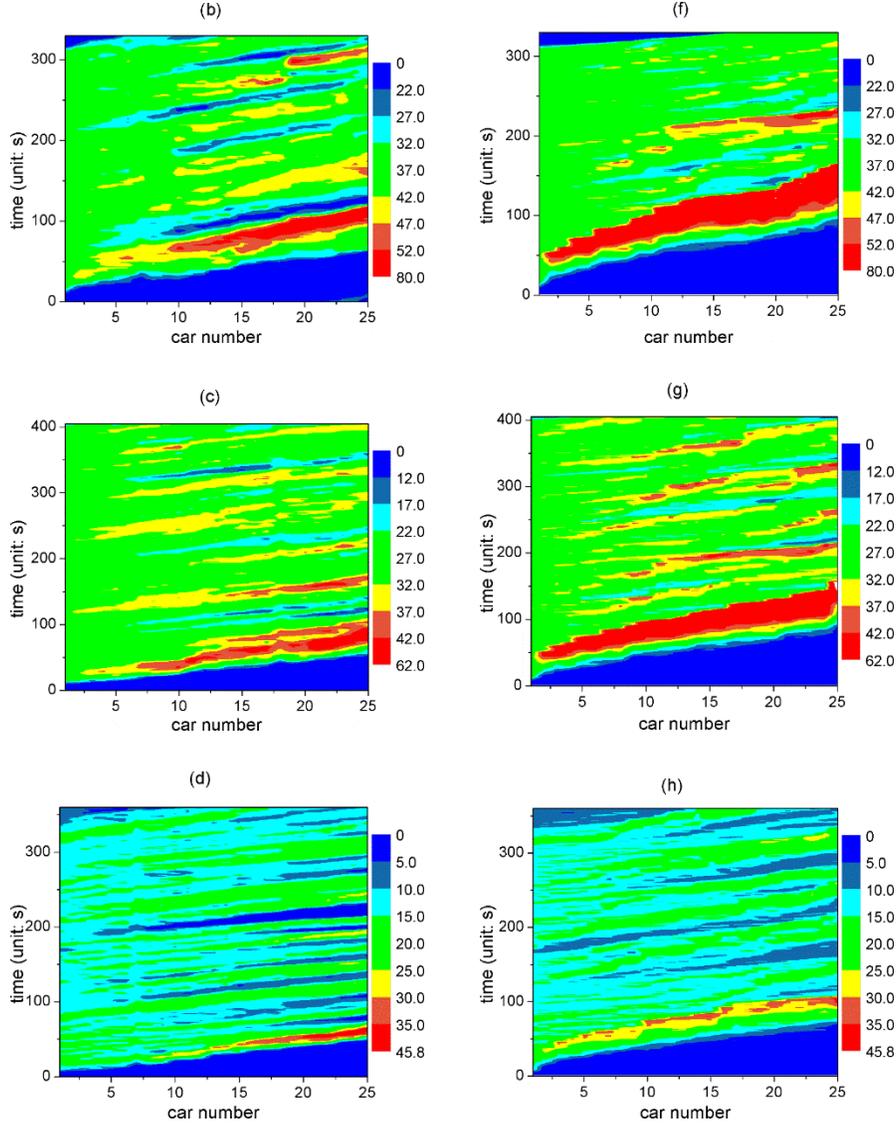

**Fig. 10.** The spatiotemporal patterns of the platoon traffic. The car speed is shown with different colors (unit: *km/h*) as function of time and car number. The left Panels show the experimental results and the right Panels show the simulation results of the DTGBLM. In the simulation, the cars are initially in a jam as in the experiment and the velocity of the leading car is set as in Fig.9.

## 5 Empirical Verification

*5.1 Calibration methodology*

Since DTGBLM contains 13 parameters, it is unwise to apply automated calibration methods directly. Some parameters can be determined by the empirical data straightforwardly. The maximum speed $v_{max}$ can be taken as the maximum of the empirical data if there are periods of FF (otherwise, it cannot be estimated). Through the average FF speed $v_{free, ave}$, the randomization probability $p_d$ is calculated by $v_{free, ave} = v_{max} - p_d$. The randomization probability $p_0$ can be estimated by $v_g \approx -(1-p_0)/k_{max}$ if the downstream propagation speed of the WMJ, i.e. $v_g$, is determined, where $k_{max}$ is the density inside the jam. Rehborn et al. (2011) have discussed several methods to measure $v_g$ and found it inside the interval [-18, -10]*km/h* while Treiber et al. (2010) give the interval [-20,-15]*km/h*. It is unnecessary to calibrate the randomization probability $p_b$, which is a robust parameter in a wide region and only related to the synchronized traffic flow. If the information offered by the empirical data is limited, we suggest that all randomization probabilities keep default values. The vehicle length $L_{veh}$ can be set as the default value $7.5m$ or determined by the trajectory data.

Furthermore, if the macroscopic traffic flow data are used, it is unnecessary to calibrate the cell length $L_{cell}$, acceleration $a_1$ and $a_2$, since the macroscopic dynamics is significantly influenced by the interaction factor $h$, the anticipation factor $g$, the randomization deceleration $d_1$ and the desire time gap $T$. If DTGBLM is calibrated by the trajectory data, it can be taken as the discrete car following model, thus the cell length $L_{cell}$ will be ignored and the anticipation factor $g$ can keep its default value that only influences value of the maximum flow rate. For the trajectory data, if the vehicle length $L_{veh}$ and the standing space gap $s_0$ are known, the vehicle length in DTGBLM should be set as their summation; if only $s_0$ is known, such as the Floating Car Data used in the following, the vehicle length $L_{veh}$ can be ignored and the standing space gap $s_0$ needs to be added into DTGBLM to calculate the space gaps, i.e. $d_f = \max(x_l − x_f − s_0, 0)$. Therefore, for the macroscopic traffic flow data, only the interaction factor $h$, anticipation factor $g$, randomization deceleration $d_1$ and desire time gap $T$ need to be calibrated by the automated methods; for the trajectory data, the maximum speed $v_{max}$, interaction factor $h$, desire time gap $T$, acceleration $a_1$ and $a_2$, randomization deceleration $d_1$ and the standing space gap $s_0$ need to be calibrated. Similar method can be easily developed for other cellular automaton models. In this section, the KKW model will be calibrated based on this method.

In this section, we firstly calibrate DTGBLM to fit the Floating Car Data used by Kesting and Treiber (2008) to test this methodology. The Genetic Algorithm will be applied for automated calibration. Then, the DTGBLM, BLM and its variants (MCD (Jiang and Wu (2003)) and ARBLM (Tian et al. (2014))) and KKW model will be calibrated and validated through the I-80 detector applied by Brockfeld et al. (2005), Wagner (2010) and Tian et al. (2014). The trial and error method will be used for automated calibration.

*5.1 Calibration by trajectory data*

For the trajectory calibration, the Floating Car Data (FCD) sets provided by Robert Bosch GmbH (DLR, 2007) are applied, which were recorded during an afternoon peak period on a fairly straight one-lane road in Stuttgart, Germany and recorded with a frequency of $10Hz$, i.e. with a time increment of $0.1s$ a new data-set has been recorded. The data set that contains both free and congested traffic is applied to calibrate DTGBLM, see Fig.11.

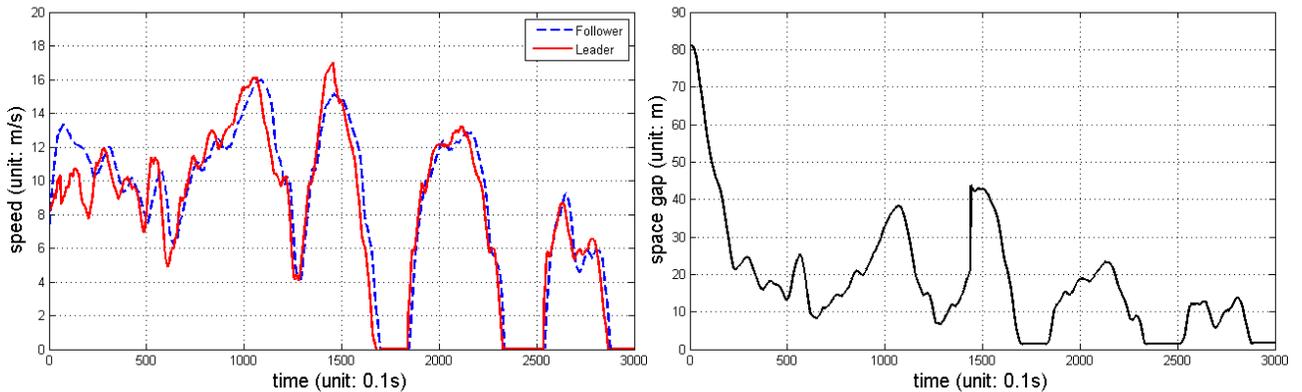

**Fig. 11.** Speed and space gap time series for the floating car data recorded by the Robert Bosch GmbH in 1995. This data set contains a passive lane change in the form that the leader of the instrumented vehicle changes to another lane/diverges off at a time of about 144s.

Initialized by the empirical gap and speed of the following vehicle: $d_f^{sim}|_{t=0} = d_f^{emp}|_{t=0}$ and $v_f^{sim}|_{t=0} = v_f^{emp}|_{t=0}$, DTGBLM is used to calculate the accelerations $a_f^{sim}(t)$, speeds $v_f^{sim}(t)$ and then the front bumper positions

$x_f^{sim}(t)$ of the following vehicle. The gaps to the leading vehicle is given by the difference between the empirical positions of the rear bumper of the leading vehicle $x_l^{emp}(t)$ and $x_f^{sim}(t)$, i.e. $d_f^{sim}(t) = \max(x_l^{emp}(t) - x_f^{sim}(t) - s_0, 0)$. Then the errors can be measured by the comparisons between $d_f^{sim}(t)$ and $d_f^{emp}(t)$. To identify whether the tested models could realistically characterize the dynamics of traffic flow, the Root Mean Squared Relative Error (*RMSRE*) to quantify the overall error is employed:

$$RMSRE = \sqrt{\frac{1}{N}\sum_i (\frac{d_{simu}^i - d_{emp}^i}{v_{emp}^i})^2} \qquad (2)$$

where $d_{simu}^i$ is the *i*th space gap of the simulation data; $d_{emp}^i$ is the *i*th space gap of the empirical data; $N$ is the number of the data points. To find the appropriate solutions to this nonlinear optimization problem, the Genetic Algorithm is applied to calibrate $v_{max}$, $h$, $T$, $a_1$, $a_2$, $d_1$, and $s_0$. Finally, the value of RMSRE is 0.242, which is consistent with the error ranges obtained in previous studies, such as, Ranjitkar et al. (2004), Brockfeld et al. (2004) and Punzo and Simonelli (2005). Fig.12 is the comparison of the simulated gaps and the empirical gaps, which shows that the variations of the real gaps can be reproduced by DTGBLM. Thus the proposed calibration method is valid for DTGBLM and it will be used to calibrate more cellular automata models by the detector data in the next section.

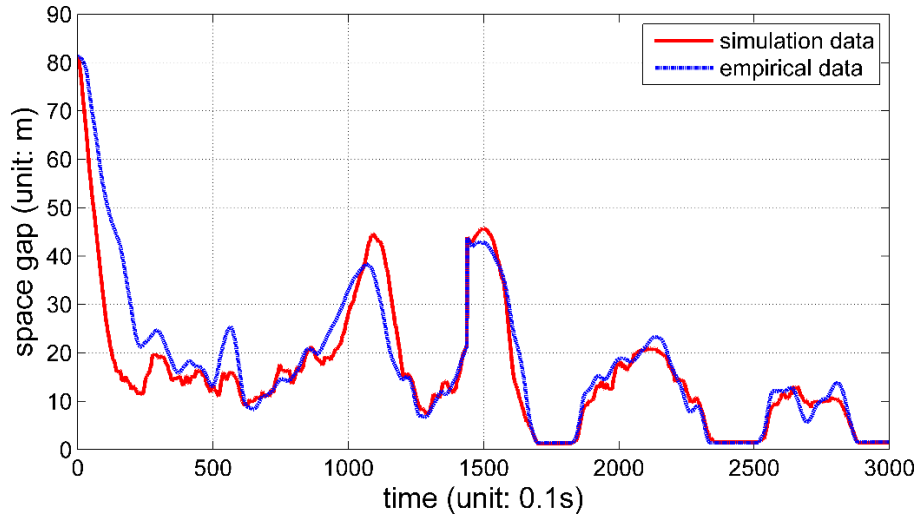

**Fig. 12.** Comparison of simulated and empirical trajectories

*5.2 Calibration and validation by detector data*

The datasets presented by NGSIM (NGSIM, 2006) are from the double loop detectors between Powell Street and Gilman Avenue on the five-lane Interstate 80 (I-80) in Emeryville, California (see Fig.13). In order to verify DTGBLM, the simulations are set as follows: two data sets from detectors 6 and 4 offer the inflow and the outflow boundary conditions, while the detector 5 measures the performance of the models by the comparisons of the simulation results at this detector with the real data. The data from detector 6 are used to drive the inflow, while the data from the detector 4 define the outflow condition.

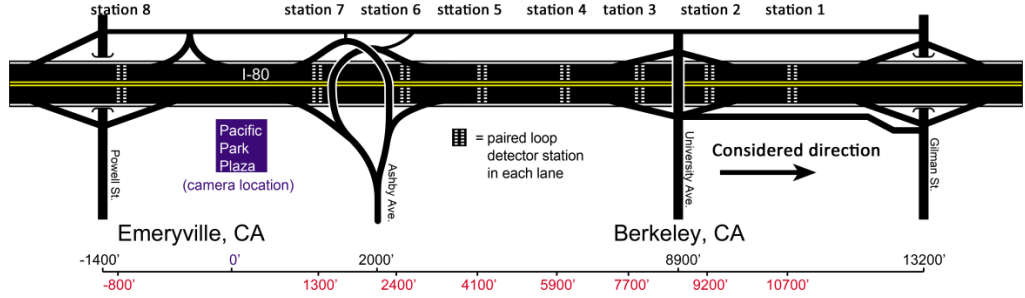

**Fig. 13.** Sketch of I-80 near Berkeley.

The data from the five-lane road are averaged into the single lane data for these detectors (see Brockfeld et al. (2005), Wagner (2010) and Tian et al. (2014)). Models will be calibrated with the data of Thursday, 07 April 2005 and then validated with the data of other five days. *RMSRE* is still used:

$$RMSRE = \sqrt{\frac{1}{N}\sum_i (\frac{v^{5i}_{simu} - v^{5i}_{ave}}{v^{5i}_{ave}})^2} \tag{3}$$

where $v^{5i}_{simu}$ is the *i*th speed of the simulation data at station 5; $v^{5i}_{ave}$ is the *i*th lane average speed of the empirical data at station 5; $N$ is the number of the data points.

The calibration and validation results are given in Table 3 and Fig.14. Due to the stochastic nature of models and simulations, separate runs of simulation with the optimal model parameters lead to different RMSRE values. Nevertheless, we found that repeat runs lead to slightly different RMSRE values. These results lead to the following conclusions: 1) although the calibration error of DTGBLM is almost the same as that of KKW, the validation errors are much smaller than that of KKW, which means that DTGBLM outperforms KKW; 2) Compared with other brake light cellular automata models, both the calibration and validation errors of DTGBLM are much more smaller. All these means that DTGBLM is the best among the five models to reproduce those detector data

**Table 3.** Calibration (07. April 2005) and validation errors (other days).

| Model | Model type | 07 Apr | 08 Apr | 11 Apr | 12 Apr | 13 Apr | 14Apr |
|---|---|---|---|---|---|---|---|
| DTGBLM | FDA | 0.217 | 0.248 | 0.179 | 0.143 | 0.153 | 0.271 |
| BLM | FDA | 0.387 | 0.500 | 0.336 | 0.312 | 0.336 | 0.469 |
| MCD | KTPT | 0.321 | 0.377 | 0.312 | 0.304 | 0.297 | 0.431 |
| ARBLM | FDA | 0.289 | 0.240 | 0.305 | 0.323 | 0.246 | 0.378 |
| KKW | KTPT | 0.223 | 0.280 | 0.192 | 0.222 | 0.201 | 0.475 |

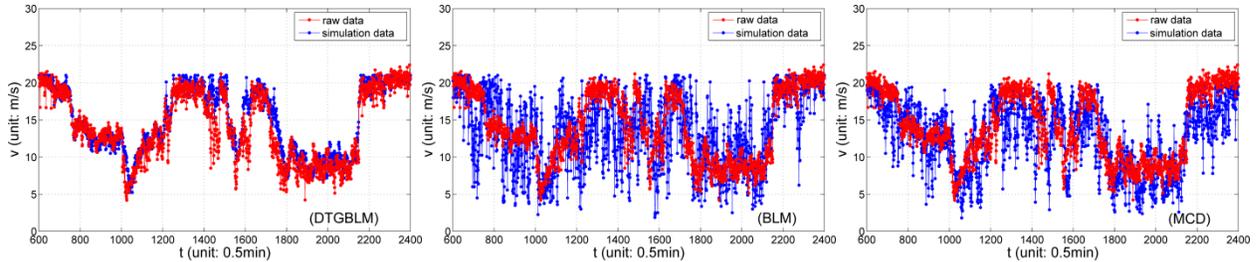

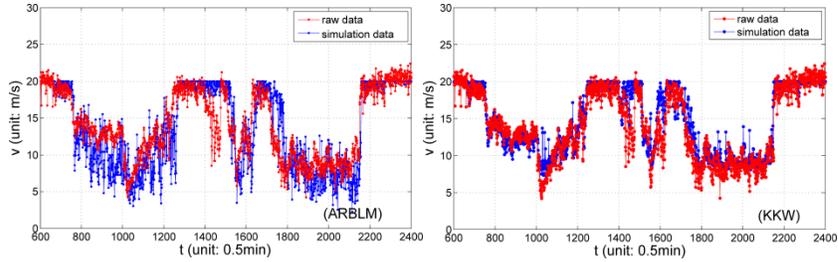
**Fig. 14.** Time series of speed at Station 5 on 07 Apr 2005.

## 6 Conclusion

There are many controversies between the two-phase traffic theory and Kerner's three-phase traffic theory. The former is based on the assumption that there is a unique relationship between the flow rate and the traffic density under the steady state condition, while the latter supposes that the hypothetical steady states of synchronized flow cover a two-dimensional region in the flow-density plane.

This paper proposes an improved cellular automaton traffic flow model based on the brake light model, in which the desired time gap of vehicles is set to be remarkably larger than one second. Although the hypothetical steady state of vehicles in the deterministic limit corresponds to a unique relationship between speeds and gaps in the proposed model, the traffic state of vehicles dynamically spans a two-dimensional region in the plane of speed versus gap, due to the various randomizations.

We have carried out simulations of the model on a circular road. It is shown that the model is able to well reproduce the free flow, synchronized flow, jams as well as the transitions among the three phases. Simulations on an open road with an on-ramp demonstrate that the model can well describe the spatiotemporal patterns of traffic flow, such as WSP and GP.

The car-following behaviors have also been studied. The evolution features of disturbances and the spatiotemporal patterns in a car-following platoon are in good agreement with the traffic experiment. Finally, the model together with other four models were calibrated and validated by the I-80 detector datasets of NGSIM, which shows that the model is the best among the five test models to reproduce those detector data.

Based on the above results, together with the findings by Jiang et al., (i.e., the growth pattern of disturbances has changed and becomes qualitatively or even quantitatively in consistent with that observed in the experiment by removing the fundamental restriction of unique velocity-spacing relationship and allowing the traffic state to span a two-dimensional region in velocity-spacing plane in two-phase models), we argue that a model can potentially reproduce the empirical and experimental features of traffic flow, provided that the traffic states are able to dynamically span a 2D speed-gap region, no matter the steady states correspond to a unique flow density relationship, or cover a 2D region in flow-density plane, or do not exist at all.

## Appendix: Abbreviations list

**Table A1.** Abbreviations list.

| Full name | Abbreviation |
|---|---|
| Free Flow | FF |
| Kerner's Three-Phase Theory | KTPT |
| Synchronized Flow | SF |
| Wide Moving Jam | WMJ |
| the transition from free flow to synchronized flow | F→S transition |
| the transition from synchronized flow to wide moving jam | S→J transition |
| the transition from free flow to wide moving jam | F→J transition |
| Oscillating Moving Jams | OMJ |
| Widening Synchronized flow Pattern | WSP |
| Dissolving General Pattern | DGP |
| General Pattern | GP |
| Widening Pinned Layer | WPL |
| Localized Synchronized flow Pattern | LSP |

**Acknowledgements:**

The authors wish to thank NGSIM and DLR for supplying the empirical data used in this article. This work is supported by the National Natural Science Foundation of China (Grant Nos. 71401120, 71271150, 71431005).